\documentclass[aps,twocolumn,superscriptaddress,floatfix]{revtex4-1}  
\usepackage{dcolumn}   
\usepackage{amsmath}
\usepackage{epsfig}
\usepackage{epstopdf}
\usepackage{graphicx}
\usepackage{color}
\usepackage{bm}        
\usepackage{amssymb}   
\usepackage{hyperref}
\usepackage[normalem]{ulem}
\usepackage[english]{babel}
\usepackage{natbib}

\begin{document}


\title{Electrostatic stability and encapsidation of charged nano-droplets}

\author{Leili \surname{Javidpour}}
\affiliation{School of Physics, Institute for Research in Fundamental Sciences (IPM), Tehran 19395-5531, Iran}
\author{An\v{z}e \surname{Lo\v{s}dorfer Bo\v{z}i\v{c}}}
\affiliation{Department of Theoretical Physics, Jo\v zef Stefan Institute, SI-1000 Ljubljana, Slovenia}
\author{Ali \surname{Naji}}\thanks{E-mail: \texttt{a.naji@ipm.ir}}
\affiliation{School of Physics, Institute for Research in Fundamental Sciences (IPM), Tehran 19395-5531, Iran}
\affiliation{Department of Applied Mathematics and Theoretical Physics, University of Cambridge, Centre for Mathematical Sciences, Cambridge CB3 0WA, United
Kingdom}
\author{Rudolf \surname{Podgornik}}
\affiliation{Department of Theoretical Physics, Jo\v zef Stefan Institute, SI-1000 Ljubljana, Slovenia}
\affiliation{Department of Physics, Faculty of Mathematics and Physics, University of Ljubljana, SI-1000 Ljubljana, Slovenia}
\affiliation{Department of Physics, University of Massachusetts, Amherst, MA 01003, USA}


\begin{abstract}
We investigate electrostatic stability of charged droplets, modeled as permeable, charged spheres, and their encapsidation
in thin, arbitrarily charged nano-shells, immersed in a neutralizing asymmetric electrolyte background. The latter consists of a small concentration of mobile multivalent counterions in a bathing solution of monovalent (positive and negative) ions. We use extensive Monte-Carlo simulations to investigate the spatial distribution of multivalent counterions and the electrostatic component of their osmotic pressure on the bounding surface of the spherical nano-shell.  The osmotic pressure can be negative (inward pressure), positive (outward pressure) or zero, depending on the system parameters such as the charge density of the droplet, the charge density of the shell, and the electrolyte screening, which thus determine the stability of the nano-container.  The counter-intuitive effects of multivalent counterions comprise the increased stability of the charged droplet with larger charge density,  increased stability in the case of encapsidating shell of charge density of the same sign as the charged droplet, as well as the possibility to dispense altogether with the encapsidating shell, its confining effect taken over by the multivalent counterions.  These dramatic effects are in stark  contrast to the conventional mean-field picture, which in particular implies that a more highly charged spherical  droplet should be electrostatically less stable because of its larger (repulsive) self-energy.
\end{abstract}


\maketitle

\section{Introduction}

For some time now viral proteinaceous capsids have been stripped off of their purely biological context and genetic cargo, and are being considered simply as {\em nano-platforms} with interesting applications in materials science and/or biomedical engineering \cite{Steinmetz2011}. Pre-assembled capsids can be used as {\em templates} for chemical engineering performed at their outer surface, their inner surface, as well as at the interface between protein capsomere subunits in order to construct multifunctional nano-reactors with catalytic walls \cite{Douglas}.  In a complementary approach, unassembled capsid protein subunits can be used in order to {\em encapsidate}, i.e.,  self-assemble a protein cage around an existing non-genetic cargo that has been suitably engineered in order to control the capsid (dis)assembly and to target the cargo delivery \cite{Aniagyei}. 

Viral capsids and virus-like particles can be furthermore used as {\em nano-scaffolds} that allow for engineered enzyme selectivity and enzyme confinement enabling the precise positioning of enzymes on nano-arrays at high resolution, unavailable without this nano-scaffolding \cite{Cardinale}. These {\em enzyme nano-carriers} based on viral capsids would thus allow us to mimick the optimized enzyme positioning otherwise accessible only within the natural biological milieu. While these nano-platforms are based on pre-assembled or co-assembled hard-shell proteinaceous containers, the chemical identity of the shell can be modified in order to control the fine-tuning of chemical reactions in self-assembled {\em nano-reactors} \cite{Vriezema}.

Vesicles and {\em soft-shell liposomes} composed of lipid or surfactant amphiphile building blocks, and block-copolymer or dendrimer {\em polymersomes}, can just as well provide the necessary scaffolding and compartmentalization for chemical engineering on the nano-scale. In the final instance, one can simply eliminate the hard and/or soft shell altogether by confining the nano-chemistry to femtoliter and subfemtoliter aqueous droplets in oil \cite{Goldner}. The properties of this {\em nano-droplet confinement}, their small size, simple manipulability, and fast mixing, can all be engineered in order to enable analytical nano-chemistry and/or enzymology at conditions that would be difficult to achieve much less to control otherwise.

While the architecture and physicochemical properties of all these nano-platforms are obviously varied, one common feature is that quite generally all these molecular nano-assemblies are composed of {\em charged molecules}. Therefore they usually respond strongly to changes in the bathing solution $p$H, its ionic strength and/or ionic composition, that together set the range of stability of these (self-)assembled molecular aggregates as quantified by the magnitude and sign of the osmotic pressure acting on the boundary of this molecular container. The response to variation in the parameters characterizing the bathing solution and the ensuing stability phase diagram are at least in part due to the intricate physicochemical environment composed of protein zwitterions, amphiphiles, monovalent counterions and added multivalent and monovalent salt ions.

In order to understand or even just to identify some of the salient features of this response, we analyze the mobile charge distribution and the emerging equilibrium of osmotic forces on the boundary of a spherical nano-container with charged (molecular) cargo, in the presence of various types of bathing ionic species. This setup presents a very sophisticated version of a {\em confined Coulomb fluid} \cite{Baus}.  We shall employ extensive Monte-Carlo (MC) simulations based on a coarse-grained model that can capture the essential thermodynamic and electrostatic aspects of this system. The enclosed cargo is modeled  as a spherical volume of charge, which we shall refer to as a ``charged nano-droplet", and may or may not be encapsidated in an arbitrarily charged, thin  ``nano-shell scaffold". These two subunits together constitute a nano-container characterized by volume as well as surface charge density (note that a charged cargo may have a surface charge density itself without the presence of  a charged encapsidating shell).  This nano-container is assumed to be permeable to and immersed in a neutralizing and asymmetric bathing electrolyte solution, consisting of a small concentration of multivalent ``counterions" (with opposite charge to the droplet's volume charge) and a monovalent salt background (with positive and negative ions).  The presence of mobile as well as fixed charges in the core or on the boundary of this nano-container defines the magnitude and sets the direction of the resultant osmotic forces acting on it surface, thus strongly influencing the stability of a nano-droplet with or without an encapsidating shell.

The most remarkable features of this system emerge when one deals with a {\em highly asymmetric} electrolyte involving  multivalent counterions with large charge valency (e.g., tri- or tetravalent ions). Multivalent counterions that are strongly electrostatically coupled to fixed charges are known to generate strong electrostatic correlations, which lie at the core of recent developments in the theory of highly charged macromolecular systems \cite{holm}. The effect of multivalent counterions is known to underlay a whole slew of counter-intuitive phenomena such as formation of large condensates of DNA \cite{Bloom2,Yoshikawa1,Yoshikawa2,Pelta} or its dense packaging in viruses and nano-capsids \cite{Plum,Raspaud, Savithri1987,deFrutos2005,Siber}, formation of large bundles of charged  polymers such as microtubules \cite{Needleman} and F-actin \cite{Angelini03,Tang} and other like-charge attraction phenomena \cite{holm,hoda_review, Naji_PhysicaA, Shklovs02,Levin02,jcp_perspective} still not completely understood in all their details. These studies have led to a major paradigm shift in our understanding of the behavior of strongly coupled Coulomb fluids from the century-old Poisson-Boltzmann (PB) framework \cite{Israelachvili,VO} to a novel strong-coupling paradigm (see Refs. \cite{holm,hoda_review,Naji_PhysicaA, Shklovs02, Levin02, jcp_perspective} and references therein) providing the conceptual background uniting all these various features.

In what follows we show that a host of counter-intuitive phenomena that go beyond  the conventional PB wisdom emerge also in the context of charged nano-containers in the presence of multivalent counterions. For instance, we show that a charged droplet with higher volume charge density can be stabilized more strongly in the presence of multivalent counterions, in stark contrast with the naive expectation that it should be destabilized by its self-repulsion. It also turns out that such a highly charged droplet can be stabilized with multivalent counterions even in the absence of any encapsidating shell, and that  its encapsidation is favored to occur more easily (i.e., with a lower free energy) with a like-charged encapsidating shell than with an oppositely charged one! These effects are accompanied by an enhanced accumulation of multivalent counterions inside the charged droplet where they form a strongly correlated structure with important repercussions for the osmotic pressure acting on the boundaries of the system.

The organization of the paper is as follows: In Section \ref{sec:mm}, we introduce our model and simulation methods. We then present our results for the distribution of multivalent ions across the charged droplet in Section \ref{sec:val} and then analyze the resultant electrostatic pressure acting on the outer surface of the droplet  (or its encapsidating shell, if present) in Section \ref{sec:res} and finally, conclude our discussion in Section \ref{sec:sum}.

\section{Model and methods}
\label{sec:mm}

\begin{figure}[t!]
\includegraphics[width=6.cm]{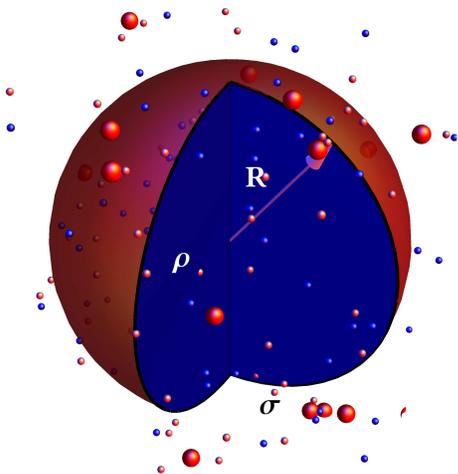}
\caption{(Color online) Schematic view of a spherical, charged nano-droplet of uniform volume charge density $\rho$ contained in a nano-shell of radius $R$ that may also have a surface charge density $\sigma$. The resulting nano-container is assumed to be permeable to mobile ions and is immersed in an asymmetric electrolyte (Coulomb fluid) comprised of monovalent and multivalent (counter-)ions (multivalent ions are shown here by bigger red spheres).}
\label{fig:sketch}
\end{figure}

Our model consists of a spherical volume of radius $R$, which is filled uniformly with a charge distribution of  volume density $\rho$. This  ``charged nano-droplet" may in general have a finite surface charge density $\sigma$ of its own, or it may be encapsidated within a thin charged ``nano-shell scaffolding" forming a charged spherical nano-container (Fig. \ref{fig:sketch}). We shall consider both bare droplets with $\sigma=0$ or encapsidated ones with a non-vanishing $\sigma$.  The spherical nano-container is considered to be  immersed in a monovalent 1:1 salt solution of bulk concentration $n_0$ and a multivalent $q$:1 salt of bulk concentration $c_0$, with $q$ being the charge valency of multivalent ions. Both the spherical shell and the charged droplet inside it are assumed to be permeable to all the mobile solution ions.
The present model can be relevant to situations encountered within the context of charged viral capsids~\cite{Baker1999,ALB2012} as well as synthetic nano-shells~\cite{Steinmetz2011,Yildiz2012} encapsidating charged bio or synthetic polyelectrolytes, within the context of nano-scaffolds confining concentrated enzyme droplets as nano-reactors \cite{Vriezema} or vehicles for targeted delivery \cite{Cardinale}, as well as in the context of compartmentalization for chemical engineering on the nano-scale \cite{Goldner}.

For typical viral nano-shells the radius varies between $5~{\mathrm{nm}}\lesssim R \lesssim50~{\mathrm{nm}}$, with a shell thicknesses usually between 1.5 and 4.5~nm, and can be much bigger in the case of nano-compartments for chemical engineering, that start as $R \gtrsim 50~{\mathrm{nm}}$. The corresponding volume charge density of the encapsidated cargo is assumed to be negative, having a value $|\rho| \lesssim 1.3$ $e_0/\mathrm{nm}^{3}$, the upper bound being typical for dense condensed DNA or other encapsidated polyelectrolytes, and the surface charge density in the range $|\sigma|\lesssim 0.4$ $e_0/\mathrm{nm}^{2}$ for virus-like nano-shells~\cite{ALB2012}.

With no loss of generality, we shall take the multivalent ions to be positively charged, $q>0$ (in most cases, we shall assume tetravalent ions with $q=4$) and thus focus on the case with $\rho<0$, but shall consider both positive and negative values for $\sigma$. Therefore, multivalent ions have opposite charge to the droplet's volume charge (but  not necessarily to the charge on its bounding surface or enclosing shell) and, hence, may be considered as  ``counterions"  belonging to the droplet. 

 The presence of both monovalent and multivalent ions makes it difficult to study the  system within a unified theoretical framework. This is because the monovalent ions are coupled weakly to fixed charges, while the multivalent ions can be generally coupled quite strongly. By using advanced statistical mechanical methodology it has been shown that the behavior of highly asymmetric mixtures with large $q>1$~\cite{Kanduc2010,Kanduc2011,Kanduc2012} can be understood by integrating out the degrees of freedom associated with monovalent ions, resulting in an approach, dubbed the  ``dressed multivalent ions" approach, with an effective screened Debye-H\"uckel (DH) interaction between all other remaining strongly-coupled charges. The corresponding  inverse Debye screening length $\kappa$ in this system depends on the bulk concentrations of both monovalent and  multivalent ions as $\kappa^2=8\pi \ell_{\mathrm{B}}(n_0+qc_0/2)$~\cite{Kanduc2011,Kanduc2012}, where $\ell_{\mathrm{B}}=\beta e_0^2/(4\pi\varepsilon\varepsilon_0)$ is the Bjerrum length with $\beta=1/k_{\mathrm{B}}T$. Even with small monovalent ions swept into the effective screening parameters, the problem remains a many-body one and because of the presence of multivalent ions can not be generally treated by usual mean-field schemes. It may however be treated using MC simulations~\cite{Kanduc2010,Kanduc2011,Kanduc2012} as we shall in this work.

In our MC simulations, we consider a large collection of mobile multivalent counterions that can permeate the encapsidated charged droplet in a bathing solution of monovalent ions, which are treated implicitly, i.e., by providing a screened DH potential between explicitly modeled multivalent ions and fixed, continuum (surface and volume) charges of the nano-container. The advantage of this model is that it enables more efficient simulations in the regime of parameters where the discrete nature of the monovalent ions and solvent molecules can be neglected. This implicit-ion approach has been tested extensively with explicit-ions MC simulations and shown to provide an accurate description for sufficiently large $q$ and sufficiently large (small) monovalent (multivalent) salt concentrations~\cite{Kanduc2010,Kanduc2011,Kanduc2012}. It breaks down outside this regime, as well as when nonlinear charge renormalization and/or Bjerrum pairing effects in the ionic mixture are important \cite{Bjerrumpairing1,Bjerrumpairing2,Bjerrumpairing3}; these latter effects however turn out to be absent or negligible in the regime of parameters that is of concern to our discussion \cite{Kanduc2010, Kanduc2011,Kanduc2012}.

The monovalent salt concentration, $n_0$, is varied in the simulations in the range between 30~mM to 300~mM and we take a small concentration $c_0$ of multivalent $q$:1 salt of the order of a few mM, in accordance with the typical values found in experiments~\cite{Savithri1987,deFrutos2005}. The ambient temperature is $T=300$~K  and the dielectric  constant of the solution is taken as that of water $\varepsilon=80$, giving $\ell_{\mathrm{B}}=0.7$~nm. We employ canonical MC simulations for dressed multivalent ions confined to a sufficiently large cubic volume with the charged nano-container (nano-droplet) located at the center and then use an iterative method in order to generate  a constant equilibrium ``bulk'' concentration of multivalent ions at large separations from the nano-container (nano-droplet), thus  simulating an open (grand-canonical) system in an efficient way.

The electrostatic part of the osmotic pressure acting on the bounding surface of the spherical nano-container (nano-droplet) is then obtained from
\begin{equation}
P=P_{DH}-\left\langle\sum_i q_i\frac{\partial \varphi_{DH}({\bf r}_i)}{\partial V}\Big|_{Q_\sigma,Q_\rho}\right\rangle,
\label{nuisgeorni}
\end{equation}
where $V=4\pi R^3/3$ is the container (droplet) volume and the partial derivatives here are to be taken at fixed values of its total surface and volume charge, i.e.,  ${Q_\sigma}=4\pi R^2 \sigma$ and  ${Q_\rho}=(4\pi/3)R^3\rho$, respectively. Also, $P_{DH}$ is the purely repulsive (positive) DH osmotic pressure that can be calculated from  the self-energy of the fixed charges (see Appendix \ref{app:DH}). The second term in Eq. (\ref{nuisgeorni}) involves contributions from the multivalent counterions and can be attractive (negative) or repulsive, and can thus play a major role in the stabilization of the charged droplet inside the shell as we show later.  The screened potential, $\varphi_{DH}({\mathbf r})$, results from the fixed charges, i.e. both the surface charge of the shell and its inner volume charge, and  can be calculated from the standard DH equation (Appendix \ref{app:DH}). The averaging is performed  over a sufficiently large number of MC steps after proper equilibration of the system. 

\begin{figure}[t!]
\includegraphics[width=8.5cm]{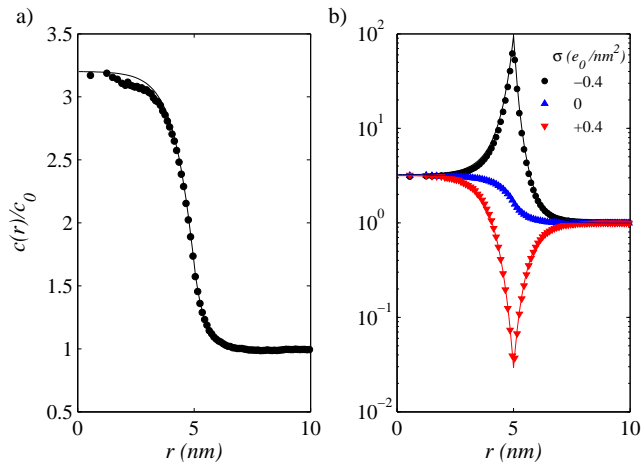}
\caption{(Color online) Density profile  of multivalent counterions with charge valency $q=4$ as a function of the radial distance from the center of a nano-container consisting of a shell of radius $R=5$~nm and surface charge density $\sigma$ encapsidating a charged droplet of volume charge density $\rho$ for a) $\rho = -0.1\,e_0/\mathrm{nm}^3$ and $\sigma = 0$, and  b) $\rho = -0.1\,e_0/\mathrm{nm}^3$ and three different values of $\sigma=-0.4, 0$ and $+0.4\,e_0/\mathrm{nm}^2$. Symbols show our  simulation data and solid lines are the limiting analytical ``dressed multivalent-ion" results, as explained in the text. The density profiles are rescaled by the bulk value $c_0=1$~mM and we have  $\kappa = 1.73\,{\mathrm{nm}}^{-1}$ for these plots.
}
\label{fig:ProfilesAgr}
\end{figure}

\section{Distribution of multivalent counterions}
\label{sec:val}

The spatial distribution of multivalent counterions can be determined from their radial (number) density profile that can be calculated straightforwardly from the output of MC simulations.  Let us first consider the situation where the nano-container is weakly charged, i.e. both $|\rho|$ and $\sigma$ are small.  In Fig. \ref{fig:ProfilesAgr}, we show the radial density profile in the case of tetravalent counterions when  $\rho = -0.1\,e_0/\mathrm{nm}^3$ and $\sigma=0$ (panel a) and compare the results for different surface charge densities on the enclosing shell (or the outer boundary of the droplet)  $\sigma=-0.4, 0$ and $+0.4\,e_0/\mathrm{nm}^2$ (panel b).  It is clear that multivalent counterions tend to accumulate inside the droplet resulting in a step-like density profile when there is no surface charge (panel a), which then falls  rapidly to the bulk concentration at large distances away from the droplet. The presence of a finite surface charge density changes the shape of the profile significantly across the shell as multivalent counterions (taken to be positively charged here) can be strongly attracted to or depleted from the interfacial region for negative or positive surface charge densities, respectively.

These results agree qualitatively with those found in the case of empty charged nano-shells \cite{leili1}, where an analytical ``dressed multivalent-ion" approach~\cite{Kanduc2010,Kanduc2011,Kanduc2012} can indeed accurately describe the simulation results in its respective regime of validity, i.e.,  when the monovalent (multivalent) salt concentration is large (small) enough and the fixed charge densities are not too large. For sufficiently large salt screening parameters, shown in  Fig. \ref{fig:ProfilesAgr}, with $\kappa = 1.73\,{\mathrm{nm}}^{-1}$, equivalent to a monovalent salt concentration of 283~mM, and tetravalent  ions concentration of 1~mM,  we find a close agreement between the simulated density profile (symbols) and the analytical  ``dressed multivalent-ion" prediction (solid lines). The latter is based on a single-particle form $c({\mathbf r}) = c_0 \exp(-\beta qe_0\varphi_{DH}({\mathbf r}))$~\cite{Kanduc2010,Kanduc2011,Kanduc2012} that follows as a leading-order result from a virial expansion of the partition function of the system in terms of the fugacity (concentration) of multivalent ions, as the latter is usually rather small in most experimental examples~\cite{Savithri1987,deFrutos2005}. This leads to a formally simple, yet  fundamentally novel approach to such asymmetric, multicomponent Coulomb systems \cite{jcp_perspective}, which combines the weak-coupling nature of monovalent ions and the strong-coupling aspects of multivalent ions in a consistent way and can thus smoothly interpolate between the usual PB or DH  limit and the standard counterion-only  {\em strong-coupling theory} introduced originally for salt-free systems~\cite{Netz,hoda_review,Shklovs02}. The above  analytical prediction will not be applicable as the volume charge density of the droplet is increased further, which is the regime of interest in this study and is, thus, accessible only via  numerical simulations.

\begin{figure}[t!]
\includegraphics[width=8cm]{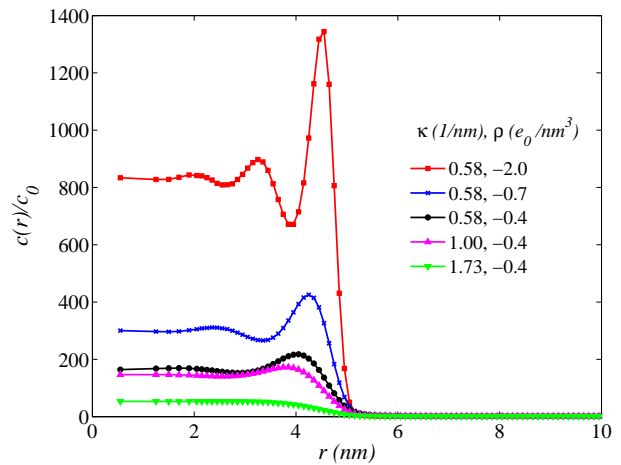}
\caption{(Color online) Rescaled density profile of multivalent counterions as a function of the radial distance from the center of a charged droplet  with $R=5$~nm and $\sigma = 0$ for different salt screening parameters $\kappa=0.58, 1$ and $1.73 \mathrm{nm}^{-1}$ and droplet charge densities $\rho = -0.4, -0.7, -2.0 \; e_0/\mathrm{nm}^3$ as shown on the graph. Here we have $q=4$, $c_0 = 1$ mM. The  pressure acting on the surface of the nano-droplet in these five different cases are $P = -1.37, -0.241, -0.077, -0.008$, and 0.120~$k_{\mathrm{B}}T/{\mathrm{nm}}^{3}$ from top to bottom.}
\label{fig:ProfilesOscil}
\end{figure}

As seen in Fig. \ref{fig:ProfilesOscil} for larger values of $|\rho|$, the density of multivalent counterions inside the nano-droplet is  significantly increased (by at least an order of magnitude) as compared with its bulk value, $c_0$. In this situation, the ion-ion interactions (repulsions) become increasingly more relevant and one can observe the formation of an ordered shell structure of multivalent counterions manifested clearly by pronounced peaks in the local density that start to develop from the outer boundary of the spherical nano-droplet inwards. Figure \ref{fig:ProfilesOscil} shows that the peak formation is enhanced as the screening parameter is decreased (from $\kappa= 1.73 \mathrm{nm}^{-1}$ down to $\kappa= 0.58  \mathrm{nm}^{-1}$ at fixed $\rho = -0.4 \; e_0/\mathrm{nm}^3$) and/or the volume charge density is increased (up to $\rho =  -2.0 \; e_0/\mathrm{nm}^3$ at fixed $\kappa= 0.58  \mathrm{nm}^{-1}$), both leading to stronger ion-ion correlations. In fact, as one can infer from the simulated pair distribution functions of multivalent ions (not shown), a distinct correlation hole is formed between individual ions inside the droplet which is indicative of a correlated liquid-like behavior. The size of the correlation hole $a$, and indeed also the distance between the peaks seen in Fig. \ref{fig:ProfilesOscil}, are consistent with a naive estimate based on the local electroneutrality  condition, giving  $a\sim (q/|\rho|)^{1/3}$.

\section{Osmotic pressure}
\label{sec:res}

\begin{figure}[t!]
\includegraphics[width=8cm]{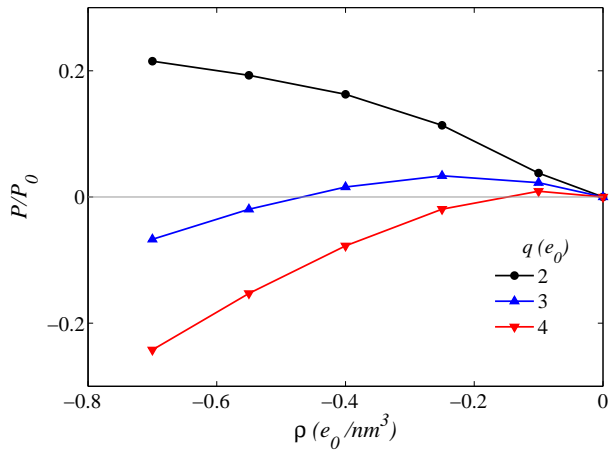}
\caption{(Color online) Electrostatic osmotic pressure acting on the nano-container (nano-droplet) bounding surface as a function of the enclosed volume charge density, calculated from MC simulations by using Eq. (\ref{nuisgeorni}),  for $R = 5$~nm, $\sigma = 0$, $\kappa=0.58 \mathrm{nm}^{-1}$, $q\,c_0 = 4$ mM and different multivalent counterion valencies $q=2, 3$ and 4. The pressure is rescaled in this plot with $P_0=1$ $k_{\mathrm{B}}T/{\mathrm{nm}}^{3} \simeq  41 ~\rm atm$.}
\label{fig:PvsDiff_q}
\end{figure}

\subsection{Charged droplet with no encapsidating shell}

As noted in Section \ref{sec:mm}, the electrostatic part of the osmotic pressure acting on the bounding surface of the charged droplet consists of two different contributions, see Eq. (\ref{nuisgeorni}). The DH self-energy of the fixed surface and volume charges is clearly positive, yielding always an {\em outward} osmotic pressure component.  The second contribution stems from multivalent counterions and can exhibit intriguing features that go beyond the standard mean-field picture usually obtained in the presence of monovalent ions only.

In this section, we focus on the special case with $\sigma=0$, corresponding to a situation where the droplet has no surface charge and/or is not encapsidated in a charged shell (the effects of a charged shell around the droplet will be studied later in more detail). Figure \ref{fig:PvsDiff_q} shows the simulated osmotic pressure as defined in Eq. (\ref{nuisgeorni}) as a function of the droplet's volume charge density $\rho$ for relatively small salt concentration with $\kappa=0.58  \mathrm{nm}^{-1}$ (equivalent to a monovalent salt concentration of 30~mM). For divalent ions, the pressure is positive and decreases in magnitude by decreasing the magnitude  of $\rho$, obviously reflecting the fact that the self-energy repulsion is still dominant. In the case of trivalent and tetravalent ions, the situation appears to be different: both show  {\em inward}  (negative) osmotic pressure for large negative droplet volume charge  density which gradually increases to zero and eventually becomes positive as $|\rho|$ is decreased  (in Fig. \ref{fig:PvsDiff_q}, we take $qc_0=4$~mM such that $\kappa$ and $n_0$ can be kept the same between different data sets). The effect is stronger for tetravalent ions, showing a larger region with negative pressures but, on the other hand, trivalent ions show an interesting non-monotonic behavior with a maximum positive osmotic pressure in the plotted region. (Note that negative pressure can be seen with divalent  ions as well but in a different range of parameters not discussed here; therefore, the exact region where  positive and negative osmotic pressure appears for a fixed $q$ depends on other parameters as well.)

A key conclusion here is that, even in the absence of a charged encapsidating shell, a spherical charged droplet can be {\em stabilized} simply by adding a small amount of multivalent counterions to the solution.

\begin{figure}[t!]
\includegraphics[width=8cm]{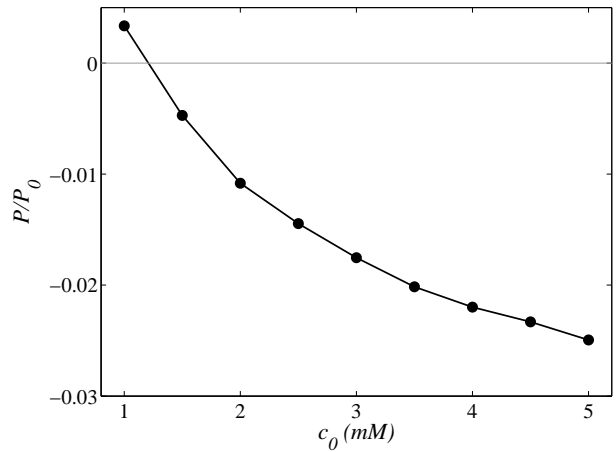}
\caption{Electrostatic osmotic pressure acting on the nano-droplet surface as a function of the bulk concentration of multivalent counterions for $R = 5$~nm, $\sigma = +0.25 \; e_0/\mathrm{nm}^2$, $\rho = -0.4 \; e_0/\mathrm{nm}^3$ $\kappa=1 \; \mathrm{nm}^{-1}$ and $q=4$.}
\label{fig:DiffC_0}
\end{figure}

\begin{figure*}[t]\begin{center}
\includegraphics[width=12cm]{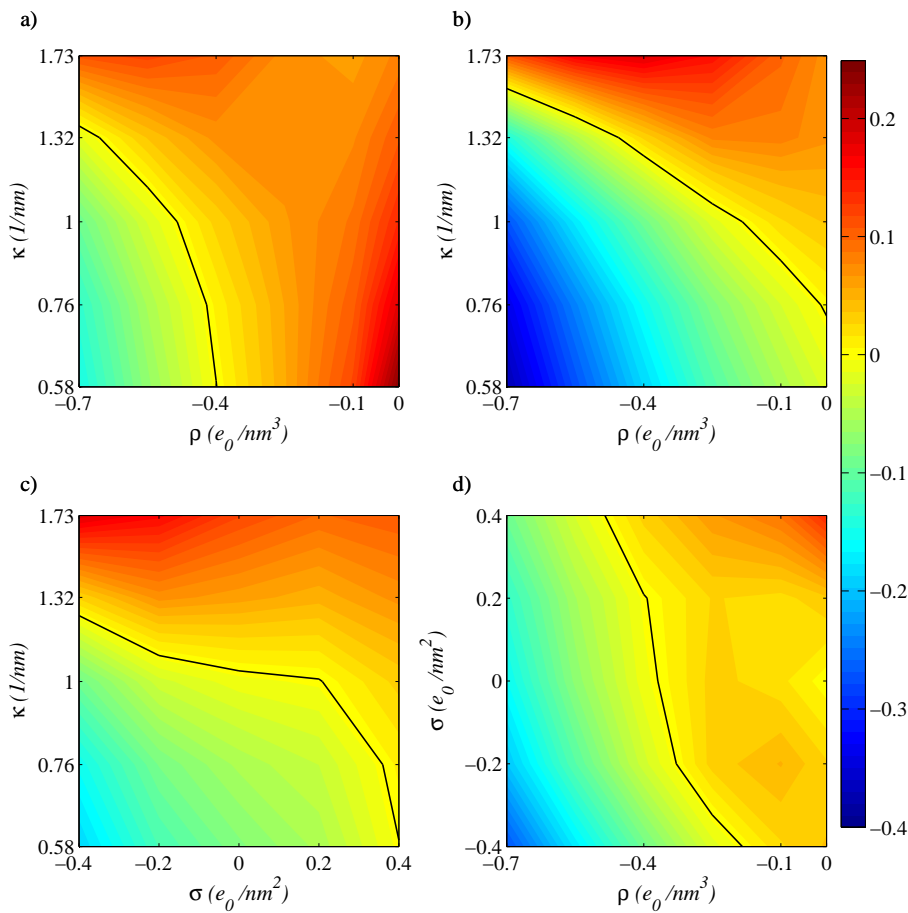}
\caption{(Color online) Phase diagrams for the rescaled (electrostatic) osmotic pressure $P/P_0$ acting on the bounding surface of a spherical nano-droplet of radius $R=5$~nm in the presence of  tetravalent counterions of bulk concentration $c_0 = 1$~mM and at fixed parameter values: a) $\sigma = +0.4 \; e_0/\mathrm{nm}^2$, b) $\sigma = -0.4 \; e_0/\mathrm{nm}^2$, c) $\rho = -0.4 \; e_0/\mathrm{nm}^3$ and d) $\kappa=1\; \mathrm{nm}^{-1}$. The solid line shows the contour line where the pressure vanishes.}
\label{fig:pd}
\end{center}\end{figure*}

Another remarkable point is that the higher the volume charge density, the stronger the negative inward osmotic pressure, and therefore the more likely it is for more highly charged droplets to be more stabilized by adding the {\em same} amount of multivalent  counterions! This counter-intuitive conclusion is in stark contrast with the mean-field wisdom which would predict a larger self-repulsion and thus a stronger {\em destabilization} when a  spherical charged droplet is more highly charged.

\subsection{Stabilization by multivalent-ion concentration}

Our results in the previous Section highlight the importance of the multivalent counterion {\em valency} in generating a stabilizing osmotic pressure on the nano-droplet. An equally important parameter in this context is  the bulk {\em concentration} of multivalent counterions, i.e., $c_0$ . As an example, we show the simulated osmotic pressure acting on a charged droplet (with $\rho = -0.4 \; e_0/\mathrm{nm}^3$) encapsidated in a positively charged shell ($\sigma = +0.25 \; e_0/\mathrm{nm}^2$) in Fig. \ref{fig:DiffC_0}. The osmotic pressure is positive below 1~mM concentration of the multivalent salt, vanishes when the concentration is slightly increased beyond this value and becomes increasingly more negative as $c_0$ is increased further.

\subsection{Encapsidation in a charged nano-shell: phase diagrams}

We now focus on the case of tetravalent counterions ($q=4$) of fixed bulk concentration of $c_0=1$~mM and investigate the role of other parameters, such as the salt screening parameter and surface charge density of the shell,  on the electrostatic stability of an encapsidated charged droplet. We summarize our results in phase diagrams such as those shown in Fig. \ref{fig:pd}, where the osmotic (electrostatic) pressure acting on the bounding surface of the droplet is color-coded with the vertical bar displaying its values in units of $P_0=1$ $k_{\mathrm{B}}T/{\mathrm{nm}}^{3} \simeq 41 ~\rm atm$.

Figures \ref{fig:pd}a and b compare two different situations: in (a) we have a negatively charged droplet encapsidated by a positively charged shell of surface charge density $\sigma = +0.4 \; e_0/\mathrm{nm}^2$, and in (b) we have the same but with a negatively charged shell of surface charge density $\sigma = -0.4 \; e_0/\mathrm{nm}^2$. These diagrams allow to compare the osmotic pressures for different values of the salt screening parameters on the vertical axes. The solid line shows the separatrix between the positive and negative values of the osmotic pressures, i.e., it shows the locus of vanishing osmotic pressure. Hence, the region below this curve corresponds to the regime where the nano-droplet is stabilized by negative osmotic pressure acting inwards, while the region above the line corresponds to the regime where the nano-droplet is destabilized mainly due to its electrostatic self-repulsion.

It is thus evident that a stronger stabilization is always achieved  by lowering the salt screening parameter. When the salt screening is strong, the system is effectively neutralized and the electrostatic correlations mediated by multivalent counterions are also strongly suppressed. Again it is obvious that highly charged droplets are more easily encapsidated by the same amount of multivalent counterions  irrespective of whether the encapsidating nano-shell bears negative or positive charge. 

The remarkable result that follows by comparing the plots in Figs. \ref{fig:pd}a and b is that a charged droplet of large volume charge density is more stable and thus {\em more easily encapsidated} by a charged nano-shell if it carries a surface charge density of the {\em same sign} as the cargo! This counter-intuitive effect is again possible only with electrostatically strongly coupled multivalent counterions.

Figure \ref{fig:pd}c displays the rescaled electrostatic osmotic pressure for a wide range of surface charge densities $|\sigma|<0.4 \; e_0/\mathrm{nm}^2$, where an inward-acting osmotic pressure on the surface of a negatively charged nano-droplet  (with $\rho = -0.4 \; e_0/\mathrm{nm}^3$) is seen for nearly the whole interval of surface charge density values, provided that $\kappa\lesssim 1$~nm$^{-1}$.

Another interesting point is that, given the fact that multivalent counterions are assumed to be positively charged,  a finite minimum amount of opposite  (negative) volume charge is needed in order to obtain a stable nano-droplet. This is shown by the contour line in  Fig. \ref{fig:pd}d. In fact, as noted before, a shell-free nano-droplet ($\sigma=0$) can be self-stabilized by the attractive pressure generated by multivalent counterions if it carries a finite volume charge density, which, evidently, has to be of opposite (negative) sign. For the parameters in the figure with $\kappa = 1\,{\mathrm{nm}}^{-1}$, this corresponds to $\rho \simeq -0.4 \; e_0/\mathrm{nm}^3$.  In the presence of a negatively (positively) charged shell, this amount will be decreased (increased) in magnitude.

\subsection{The role of  nano-droplet size}

\begin{figure}[t!]
\includegraphics[width=8cm]{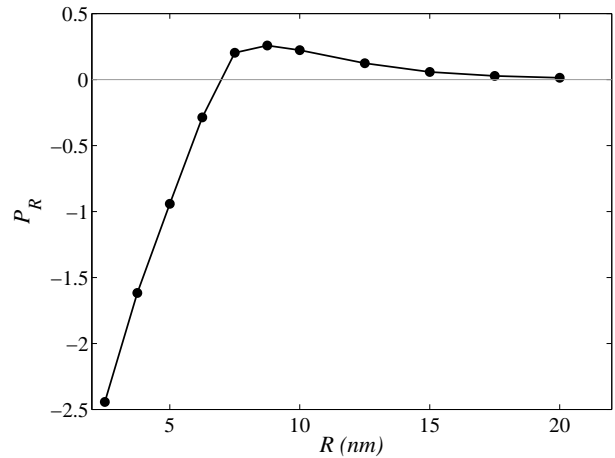}
\caption{Rescaled (electrostatic) osmotic pressure $P_{R} = {\rm sgn}(P') \times \log_{10}{(1+100~|P'|)}$ (where $P' = P/P_0$ with $P_0=1$~$k_{\mathrm{B}}T/{\mathrm{nm}}^{3}$)  from MC simulations plotted as a function of the nano-droplet radius $R$  for $q = 4$, $c_0=1$~mM, $\kappa = 0.58\,{\mathrm{nm}}^{-1}$, $\sigma = 0$, and a constant droplet charge of  $Q_{\rho}=-209\; e_0$ (equivalent to a volume charge density of $\rho = -0.4\,e_0/\mathrm{nm}^3$ if we take $R = 5$~nm). The rescaling of the osmotic pressure is done in such a way that its negative and positive values can be shown on the same log-linear scale and points of zero pressure are correctly represented.}
\label{fig:ConstQ_Cps}
\end{figure}

It is important to note that the volume charge density of the nano-droplet, and not its total charge, is the key physical factor that can determine the electrostatic component of the osmotic pressure in the presence of multivalent counterions. This is because multivalent counterions couple strongly to the local electrostatic potential when inside the droplet. Figure~\ref{fig:ConstQ_Cps} shows that for a constant total droplet charge with $Q_{\rho}=-209\; e_0$ and  $Q_{\sigma}=0$, the negative osmotic pressure is significantly reduced in magnitude when the droplet size is only modestly increased; i.e.,  it increases from around $P\simeq -113$~atm at $R\simeq 2.5$~nm and vanishes at $R_c\simeq 7$~nm and then turns positive for larger droplets. The (electrostatic) osmotic pressure exhibits an interesting {\em non-monotonic} behavior with a maximum positive value which gradually falls off to zero again when $R$ is increased further.

The above behavior thus indicates that a charged droplet of fixed total charge  can collapse in the presence of multivalent counterions to a smaller size (unless stabilized by other non-electrostatic forces) if it crosses a ``potential barrier", which is determined by the value of the radius, $R_c$, corresponding to a vanishing osmotic pressure. Conversely, the droplet can expand and dilute away in the solution if it is larger in size than this threshold value.

\section{Conclusion and discussion}
\label{sec:sum}

We investigated the effects of multivalent counterions on the stability of a permeable, spherical charged nano-droplet with and without an encapsidating charged shell, immersed in an asymmetric electrolyte, consisting of a weakly coupled monovalent and a strongly coupled multivalent salt. The former has a bulk concentration of the order of the physiological salt concentration ($\sim 100$~mM),  while the latter does not exceed a few mM.  Extensive MC simulations within a primitive model, including only the most essential features of the system, enable us to elucidate several quite unusual and unexpected consequences of electrostatic interactions that emerge because of the presence of strongly coupled multivalent counterions.

For sufficiently highly charged droplets, multivalent counterions are strongly accumulated inside the droplet (with densities much larger than their bulk concentration) and form a layered structure exhibiting a series of pronounced peaks in their radial density profile. In this respect, the highly charged droplets with multivalent counterions are similar to spherically confined plasmas, usually referred to as {\em Yukawa balls}~\cite{Bonitz2010,Totsuji2005,Henning2007}. In this case, too, a mean-field approximation ceases to give an adequate description of the system even for moderate couplings where one observes the emergence of an outer shell~\cite{PRE-YukPlasma}. With stronger couplings and increased screening, the deviation from the mean-field limit increases leading to a formation of even more pronounced multiple shell structure~\cite{PRL-YukPlasma}. At zero temperature, the charges then form narrow concentric shells and their properties can be obtained from simpler onion-shell models~\cite{Kraeft2006,Hasse1991, Tsuruta1993, Totsuji2005, Henning2007}. The charges in these shells exhibit imperfect hexagonal positional order due to the incompatibility between a perfect lattice and a curved surface and due to the incommensurability of two adjacent shells~\cite{Hasse1991}. When approaching large clusters, the competition between bulk order and spherical order becomes more pronounced. Thus, one expects the shell formation to vanish and a regular volume order of a bcc or fcc lattice to prevail in the core. Similar scenarios can be expected also for highly charged droplets with multivalent counterions but in the part of the parameter space that is unrealistic for the type of charged nano-containers and nano-shells that we investigate here.

In the regime of incipient structurization of the multivalent counterions inside the nano-droplet, we also find that the osmotic pressure acting on the bounding surface of the droplet (or its enclosing shell) can exhibit negative values, i.e., the osmotic force is directed inwards into the droplet, stabilizing it against breakdown. We have shown that multivalent counterions are the reason for this anomalous behavior and thus a large volume charge density of the droplet coupled to the multivalent counterions can lead to its stabilization, even in the absence of enclosing shell. This is certainly contrary to the mean-field wisdom \cite{Lebovka} that would predict a destabilization of a charged droplet due to its electrostatic self-repulsion as ascertained in seminal works of Rayleigh \cite{Rayleigh} and later of Weizs\" acker \cite{Weizsacker}. Our results on the stability of charged spherical aggregates in the presence of strongly coupled (multivalent) counterions show that no additional  short-range attractive interaction is needed to stabilize charged aggregates \cite{Lebovka}. The attractive (electrostatic) osmotic pressure component, originating in the mobile multivalent counterions, is enough to stabilize the charged droplet.  

Another remarkable finding is that a charged droplet is more easily encapsidated (exhibiting a lower free energy) if its volume charge is more densely packed, provided that only a small amount of multivalent counterions is present in the system, and that a more stable encapsidation is achieved by an enclosing shell whose surface charge has the same sign as the charged droplet itself. This effect, too, is connected with the strong electrostatic interactions mediated by the multivalent ions and can not be rationalized within a mean-field-type argumentation. We will present a more detailed analysis of the stabilization phenomenon elsewhere.

The main lesson of our study is that the presence of an electrostatically strongly coupled component in the bathing electrolyte solution can fundamentally change the stability properties of charged aggregates introducing counter-intuitive parameter dependencies that can not be understood within the confines of the standard mean-field paradigm.

\begin{acknowledgments}
A.N. acknowledges support from the Royal Society, the Royal Academy of Engineering, and the British Academy. A.L.B. and R.P. acknowledge support from the ARRS Grants No. P1-0055 and J1-4297.
\end{acknowledgments}

\appendix

\section{Calculation of $\varphi_{DH}$ and $P_{DH}$}
\label{app:DH}

The screened potential, $\varphi_{DH}({\mathbf r})$, results from the fixed charges, i.e. both the surface charge of the nano-shell and the inner volume charge of the nano-droplet, and
can be
calculated from the standard DH equation.
The potential for the exterior region, $\varphi_>=\varphi_{DH}(r\geqslant R)$,  must satisfy the DH equation,
\begin{equation}
(\nabla^2-\kappa^2)\varphi_>=0,
\end{equation}
but the potential inside the nano-container (nano-droplet), $\varphi_<=\varphi_{DH}(r\leqslant R)$, has to satisfy the non-homogeneous DH equation,
\begin{equation}
(\nabla^2-\kappa^2)\varphi_<=-\frac{\rho}{\epsilon\epsilon_0}.
\end{equation}
The boundary conditions are as follows: $\varphi_<$ has to be finite at the origin, whereas $\varphi_>$ has to go to zero at infinity. Both potentials must take
on the same value at $r=R$, and their derivatives must satisfy
\begin{equation}
\frac{\partial\varphi_<}{\partial r}-\frac{\partial\varphi_>}{\partial r}\Bigg|_{r=R}=\frac{\sigma}{\epsilon\epsilon_0}.
\end{equation}
The solution to these equations is obtained as
\begin{equation}
\varphi_<(r)=\frac{\sinh\kappa r}{r}\frac{\sigma\kappa^2R-\rho(1+\kappa R)}{\kappa^3\epsilon\epsilon_0(\cosh\kappa R+\sinh\kappa
R)}+\frac{\rho}{\kappa^2\epsilon\epsilon_0},
\end{equation}
and
\begin{equation}
\varphi_>(r)=\frac{e^{-\kappa(r-R)}}{r}\left(\frac{\rho R}{\kappa^2\epsilon\epsilon_0}+\frac{\sigma\kappa^2R-\rho(1+\kappa
R)}{\kappa^3\epsilon\epsilon_0(1+\coth\kappa R)}\right).
\end{equation}
Hence,  the DH self-energy of the fixed charges can be expressed as
\begin{equation}
{\mathcal F}_{DH}=\frac{4\pi}{2}\int_0^R\rho\varphi_<(r)r^2\mathrm{d}r+\frac{Q_{\sigma}\varphi(R)}{2},
\end{equation}
where $Q_{\sigma}=4\pi R^2\sigma$
is the total surface charge of the enclosing shell, and
\begin{eqnarray}
\nonumber \varphi(R)&=&\frac{\sigma}{\kappa\epsilon\epsilon_0(1+\coth\kappa R)}\\
&+&\frac{\rho}{\kappa^2\epsilon\epsilon_0}\left(1-\frac{1+\kappa R}{\kappa R(1+\coth\kappa R)}\right).
\end{eqnarray}
Thus the full DH expression for ${\mathcal F}_{DH}$ follows as
\begin{eqnarray}
\nonumber {\mathcal F}_{DH}&=&\frac{2\pi R^2\sigma^2}{\kappa\epsilon\epsilon_0(1+\coth\kappa R)}\\
\nonumber &+&\frac{2\pi R^2\sigma\rho}{\kappa^2\epsilon\epsilon_0}\left(1-\frac{1+\kappa R}{\kappa R(1+\coth\kappa R)}\right)\\
\nonumber &+&\frac{2\pi R^3\rho^2}{3\kappa^2\epsilon\epsilon_0}\\
&+&2\pi\rho\frac{\sigma\kappa^2R-\rho(1+\kappa R)}{\kappa^3\epsilon\epsilon_0(1+\tanh\kappa R)}\frac{\kappa R-\tanh\kappa R}{\kappa^2}.
\end{eqnarray}
The electrostatic (osmotic) pressure acting on the bounding surface of the spherical nano-container (nano-droplet) can be obtained from  $P_{DH}= - \partial {\mathcal F}_{DH}/{\partial V}$,
where $V=4\pi R^3/3$ is the container (droplet) volume.



\begin{thebibliography}{99}

\bibitem{Steinmetz2011}
N. Steinmetz and M. Manchester,
{\em Viral Nanoparticles: Tools for Materials Science and Biomedicine} (Pan Stanford Publishing, 2011).

\bibitem{Douglas}
T. Douglas and M. Young, Science {\bf 312}, 873 (2006). 

\bibitem{Aniagyei}
S. E. Aniagyei, C. DuFort, C. Cheng Kao and B. Dragnea, J. Mater. Chem. {\bf 18}, 3763 (2008). 

\bibitem{Cardinale}
D. Cardinale, N. Carette and T. Michon, Trends in Biotechnology {\bf 30}, 369 (2012). 

\bibitem{Vriezema}
D. M. Vriezema, M. Comellas Aragones, J. A. A. W. Elemans, J. J. L. M. Cornelissen, A. E. Rowan, and R. J. M. Nolte, Chem. Rev. {\bf 105}, 1445 (2005). 

\bibitem{Goldner}
L. S. Goldner, A. M. Jofre, and Jianyong Tang, Methods in Enzymology {\bf 472},  61 (2010).

\bibitem{Baus}
M. Baus, J. Hansen, Phys.\ Rep. {\bf 59},  1  (1980).

\bibitem{holm} 
C. Holm, P. Kekicheff,  R. Podgornik (Eds.), {\em Electrostatic Effects in Soft Matter and Biophysics} (Kluwer Academic, Dordrecht, 2001).

\bibitem{Bloom2}
V.A.  Bloomfield, Curr. Opin. Struct. Biol. {\bf 6}, 334 (1996). 

\bibitem{Yoshikawa1} K. Yoshikawa, Adv. Drug Deliv. Rev. {\bf 52}, 235 (2001).

\bibitem{Yoshikawa2}  M. Takahashi, K. Yoshikawa, V.V. Vasilevskaya, A.R. Khokhlov, J. Phys. Chem. B {\bf 101},
9396 (1997).

\bibitem{Pelta} J. Pelta, D. Durand, J. Doucet, and F. Livolant, Biophys. J. {\bf 71}, 48 (1996). J. Pelta, F. Livolant, and J.-L. Sikorav, J. Biol. Chem. {\bf 271}, 5656 (1996).

\bibitem{Plum} G. E. Plum and V. A. Bloomfield, Biopolymers {\bf 27}, 1045 (1988).

\bibitem{Raspaud} E. Raspaud, I. Chaperon, A. Leforestier, and F. Livolant. Biophys. J. {\bf 77}, 1547 (1999).

\bibitem{Savithri1987} 
H. S. Savithri, S. K. Munshi, S. Suryanarayana, S. Divakar and M. R. N. Murthy,
J. Gen. Virol. {\bf 68}, 1533 (1987).

\bibitem{deFrutos2005} 
M. {de Frutos}, S. Brasiles, P. Tavares and E. Raspaud,
Eur. Phys. J. E {\bf 17}, 429 (2005).

\bibitem{Siber} A. Leforestier, A. Siber, F. Livolant and R. Podgornik, Biophys. J. {\bf 100}, 2209 (2011).

\bibitem{Needleman}
D. J. Needleman, M. A. Ojeda-Lopez, U. Raviv, H. P. Miller, L. Wilson, C. R. Safinya, Proc. Natl. Acad. Sci. USA {\bf 101}, 16099 (2004).

\bibitem{Angelini03}
T.E. Angelini, H. Liang, W. Wriggers, G.C.L. Wong, Proc. Natl. Acad. Sci. USA {\bf 100},
8634 (2003).

\bibitem{Tang} 
J.X. Tang, T. Ito, T. Tao, 
P. Traub, P.A. Janmey, Biochemistry {\bf 36}, 12600 (1997). 

\bibitem{hoda_review}
H. Boroudjerdi, Y.W. Kim, A. Naji, R.R. Netz, X. Schlagberger and A. Serr, Phys.~Rep. {\bf 416}, 129 (2005).

\bibitem{Naji_PhysicaA}
A. Naji, S. Jungblut,  A.G. Moreira,  R.R. Netz, Physica A {\bf 352}, 131 (2005). 

\bibitem{Shklovs02}  A. Yu. Grosberg, T. T. Nguyen,  B. I. Shklovskii, 
Rev. Mod. Phys. {\bf 74}, 329 (2002).

\bibitem{Levin02} 
Y. Levin, Rep. Prog. Phys. {\bf 65}, 1577 (2002). 

\bibitem{jcp_perspective}
A. Naji, M. Kandu\v c, J. Forsman, R. Podgornik, ``Perspective: Coulomb fluids -- weak coupling, strong coupling, in between and beyond", submitted to J. Chem. Phys. (2013); e-print: arXiv:1307.5130

\bibitem{Israelachvili}
J. Israelachvili, {\em Intermolecular and Surface Forces} (Academic Press, London, 1991). 

\bibitem{VO} 
E.J. Verwey, J.T.G. Overbeek, {\em Theory of the Stability of Lyophobic Colloids} (Elsevier, Amsterdam, 1948).

\bibitem{ALB2012}
A. {Lo\v{s}dorfer Bo\v{z}i\v{c}}, A. \v{S}iber and R. Podgornik,
J. Biol. Phys. {\bf 38}, 657 (2012).

\bibitem{Baker1999}
T. S. Baker, N. H. Olson and S. D. Fuller,
Microbiol. Mol. Biol. Rev. {\bf 63}, 862 (1999).

\bibitem{Yildiz2012}
I. Yildiz, I. Tsvetkova, A. M. Wen, S. Shukla, H. Masarapu, B. Dragnea and N. Steinmetz,
RSC Advances {\bf 2}, 3670 (2012).

\bibitem{Kanduc2010}
M. Kandu\v{c}, A. Naji, J. Forsman and R. Podgornik,
J. Chem. Phys. {\bf 132}, 124701 (2010).

\bibitem{Kanduc2011}
M. Kandu\v{c}, A. Naji, J. Forsman and R. Podgornik, Phys. Rev. E {\bf 84}, 011502 (2011).

\bibitem{Kanduc2012}
M. Kandu\v{c}, A. Naji, J. Forsman and R. Podgornik, J. Chem. Phys. {\bf 137}, 174704 (2012).

\bibitem{Bjerrumpairing1}
N. Bjerrum, Kgl. Danske Videnskab. Selskab. Mat.-fys. Medd.  {\bf 7}, 1 (1926).

\bibitem{Bjerrumpairing2}
J. Zwanikken, R. van Roij, J. Phys.: Condens. Matter  {\bf 21}, 424102  (2009).

\bibitem{Bjerrumpairing3}
M. E. Fisher, Y. Levin, Phys. Rev. Lett. {\bf 71}, 3826 (1993).

\bibitem{leili1}
L. Javidpour, A. {Lo\v{s}dorfer Bo\v{z}i\v{c}}, A. Naji, R. Podgornik, submitted to J. Chem. Phys. (2013); e-print: arXiv:1212.0731

\bibitem{Netz}
R. R. Netz, Euro. Phys. J. E {\bf 5}, 557 (2001); A. G. Moreira and R. R. Netz, {\em ibid} {\bf 8}, 33 (2002).


\bibitem{Bonitz2010}
M. Bonitz, C. Hennig, and D. Block, Rep. Prog. Phys. {\bf 73}, 066501 (2010).

\bibitem{Totsuji2005}
H. Totsuji, T. Ogawa, C. Totsuji, and K. Tsuruta, Phys. Rev. E {\bf 72}, 036406 (2005).

\bibitem{Henning2007}
C. Henning, P. Ludwig, A. Filinov, A. Piel, and M. Bonitz, Phys. Rev. E {\bf 76}, 036404 (2007).

\bibitem{PRE-YukPlasma}
H. Bruhn, H. K\"ahlert, T. Ott, and M. Bonitz, Phys. Rev. E {\bf 84}, 046407 (2011).

\bibitem{PRL-YukPlasma}
M. Bonitz, D. Block, O. Arp, V. Golubnychiy, H. Baumgartner,
P. Ludwig, A. Piel, and A. Filinov, Phys. Rev. Lett. {\bf 96}, 075001 (2006).

\bibitem{Hasse1991}
R. W. Hasse and V. V. Avilov, Phys. Rev. A {\bf 44}, 4506 (1991).

\bibitem{Tsuruta1993}
K. Tsuruta and S. Ichimaru, Phys. Rev. A {\bf 48}, 1339 (1993).

\bibitem{Kraeft2006}
W. D. Kraeft and M. Bonitz, J. Phys.: Conf. Ser. {\bf 35}, 94 (2006).

\bibitem{Lebovka} N.I. Lebovka, ``Aggregation of Charged Colloidal Particles", Adv. Polym. Sci. (2012); doi: 10.1007/12\underline{ }2012\underline{ }171

\bibitem{Rayleigh} Lord Rayleigh (W. Strutt),  Philos. Mag. {\bf 14}, 184 (1882). 

\bibitem{Weizsacker} C.F. von Weizs\" acker, Z. Phys. {\bf 96}, 431 (1935).

\end{thebibliography}
\end{document}